\documentclass[preprint,info]{epl}
\usepackage{epsf}

\title{Dynamic scaling and universality in evolution\\ 
of fluctuating random networks}

\author{Miroslav Kotrla\inst{1}
\thanks{E-mail: \email{kotrla@fzu.cz}}
\and Franti\v{s}ek Slanina\inst{1}
\and Jakub Steiner\inst{2}}
\shortauthor{Miroslav Kotrla \etal}
\shorttitle{Dynamic scaling of fluctuating networks}
\institute{
     \inst{1}Institute of Physics, Academy of Sciences 
	of the Czech Republic, Na Slovance 2, \\
	182~21 Prague 8, Czech Republic \\
     \inst{2} Charles University at Prague, Department 
	of Theoretical Physics,
	V Hole\v{s}ovi\v{c}k\'ach 2, 
	180~00 Prague 8, Czech Republic
}


\begin{document}

\maketitle

\begin{abstract}
We found that models of evolving random networks 
exhibit dynamic scaling similar to scaling of growing surfaces.
It is demonstrated by numerical simulations of two variants of
the model in which nodes are added as well as removed
{[}{\it Phys. Rev. Lett.} {\bf 83}, 5587 (1999){]}.
The averaged size and connectivity of the network 
increase as power-laws in early times 
but later saturate.
Saturated values and times of saturation change with paramaters
controlling  the local evolution of the network topology.
Both saturated values and times of saturation
obey also  power-law dependences on controlling parameters.
Scaling exponents are calculated and universal features are discussed.
\end{abstract}

\section{Introduction}
In a couple of last years, there has been an increasing activity 
in the study of
the structure and the time evolution of complex random 
networks  \cite{review1,review2,review3}. 
Real data have been analyzed and at the same time
many simplified models have been formulated and investigated. 
The applications in different contexts are
ranging from evolution of WWW and Internet, over
evolution of metabolic networks to the structure of  social networks,
or linguistic networks.
Detailed mechanisms of evolution  differ from one problem 
to another, nevertheless,  some common 
features have been already identified.
For example, it was revealed that many networks are
{\it scale free}, i.e., the distribution of the connectivity 
of nodes (vertices) is a power-law. 
It was shown that the
scale free structure can be reproduced 
in the case of growing systems by models
with the preferential attachment for adding new nodes \cite{prefAtt}.

The challenging open problem is to clarify 
what are the universal feature of network dynamics, if any.
The concept of scale free networks does not provide
a classification into universality classes. A wide range of exponents 
was found by analysis of the real data,
and the exponents in models with preferential attachment
can take an arbitrary value larger than two 
\cite{dorogovtsev00,krapivsky00}.
Moreover, this concept deals with the internal structure
of the networks rather than with possible types of dynamics.

It is of interest to explore the time dependence of global
characteristics of networks.
Two such global quantities are the size of the network $N(t)$
and the mean connectivity $\overline{k}(t)$.
The size of the network is the total number of nodes and the mean
connectivity is defined as the average of the degrees 
of nodes (the local connectivities) over all nodes in the network.
Different types of the network evolution can be encountered.
In some cases, one observe a rapid increase of the network size,
the typical example being the Internet.
Another systems, e.g. linguistic networks, 
seem to be in a stationary state with an approximately constant size.
Similarly the mean connectivity can also increase or saturate.
The situation clearly depends on the time scale considered.
A network which is rapidly growing in a certain time interval  may 
after some transient reach a stationary state.
A system which looks 
to be stationary or only slowly varying
in a  limited time interval may exhibit strong fluctuations
on a long time scale.

The network size and the mean connectivity are
expected to fluctuate in real systems.
The fluctuation and/or shortage of reliable real data in a
sufficiently long time interval can result 
in difficulty or even impossibility of tracing the 
general trends. 
However, the fluctuations can be reduced 
by averaging over many independent realizations
when the evolution is explored by the simulations of 
simplified  models.
Then the information about the time
behaviour can be extracted, and bias or saturation can be revealed.
One can ask what are the time dependences of the global 
quantities,
whether there are generic scenarios of the network evolution etc.

In this Letter, we study the time evolution of global characteristics
in  the model which has been recently introduced by us in the context 
of study of the biological evolution \cite{sla_ko_99,sla_ko_00}.
We found that the averaged network size and the averaged
mean connectivity
initially increase as  power-laws and later saturate.
We show that these   quantities
exhibit dynamical scaling
analogous to the scaling revealed previously in the study of surface growth
\cite{fam_vic_85}.
The existence of scaling invariance suggests the possible
universal behaviour. Therefore, we check the sensitivity to
modifications of the model.

\section{Model}
To fix ideas, we briefly describe our model.
The model is based on extremal dynamics \cite{ba_sne_93}
with particular rules
for adding and removing a node.
The system is composed of $N$ nodes, connected by links.
The state of each node is described
by a single dynamical variable $b\,\epsilon\, (0,1)$, called barrier. 
In each  step, the following dynamical rules \cite{sla_ko_99}
are applied:

{\it (i) } The node with minimum $b$ is found and mutated. 
The barrier of the mutated node is replaced by a new random
value $b^\prime$. 

{\it (ii) } 
The barriers of all nodes linked to 
the selected unit are also replaced
by new random numbers. 
If $b^\prime$ is larger than barriers of all its linked neighbours,
the node gives birth to a new node (speciation). If $b^\prime$ 
is lower than barriers of all linked neighbours, the 
node dies out (extinction). In remaining cases, 
the number of nodes remains unchanged.

The motivation of these rules is the following.
Well-adapted nodes 
proliferate more rapidly and chance for speciation is higher. 
However, if the local diversity, measured by connectivity of the node, is
bigger, there are fewer available possibilities for the change
and the probability of speciation is lower. On the other hand,
poorly adapted nodes are more vulnerable to extinction, but at the
same time larger diversity (larger connectivity) may favour the survival.

{\it (iii) } Speciation means, that a new node is added
into the system, with a random barrier. 

{\it (iv) } The links are established between the new node and the 
neighbours of the mutated node:  each link of the ``mother'' node is
inherited with the probability $p$ by the ``daughter'' node. 
The rule {\it (iv) } reflects the fact that the new node is 
to a certain extent 
a copy of the 
original, so the relations to the environment will be initially
similar to the ones the old node has. 
Moreover, if a node which speciates has only one neighbour, a link between
``mother'' and ``daughter'' is also established.   
Similar models with the duplication of the local geometry
were recently used
for modeling of  protein interaction network \cite{vazquez01,sole,kim}.

{\it (v) }  
When the node is extinct it means, that the node is removed
without any substitution and all links it has, are broken.
As a boundary condition, we use the
following exception: if the network
consists of a single isolated node only, it never dies out.
We suppose that a node completely isolated
from the rest of the system has lower chance to survive. This
leads to the following rule.

{\it (vi) } 
If a node dies out, all its neighbours which
are not connected to any other node die out
with the probability $p_{\rm sing}$. 
We call this kind of
extinctions {\it singular extinctions}.

\begin{figure}
\twofigures[scale=0.7]{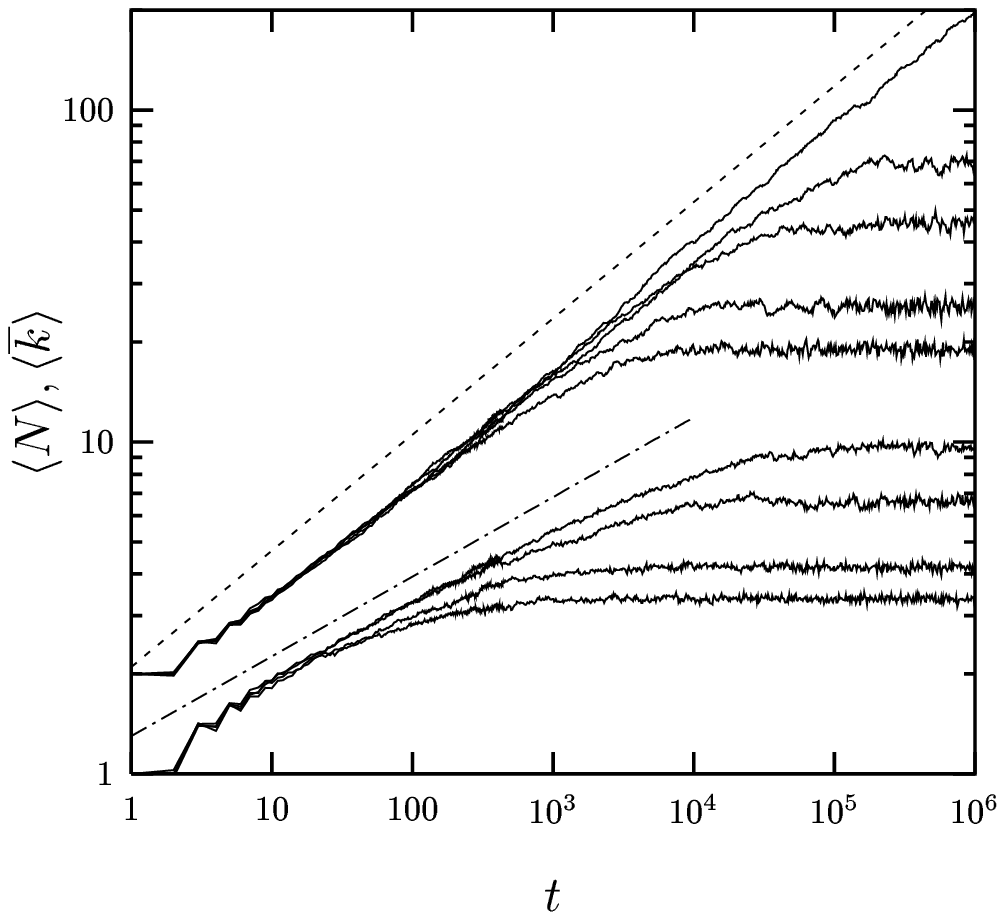}{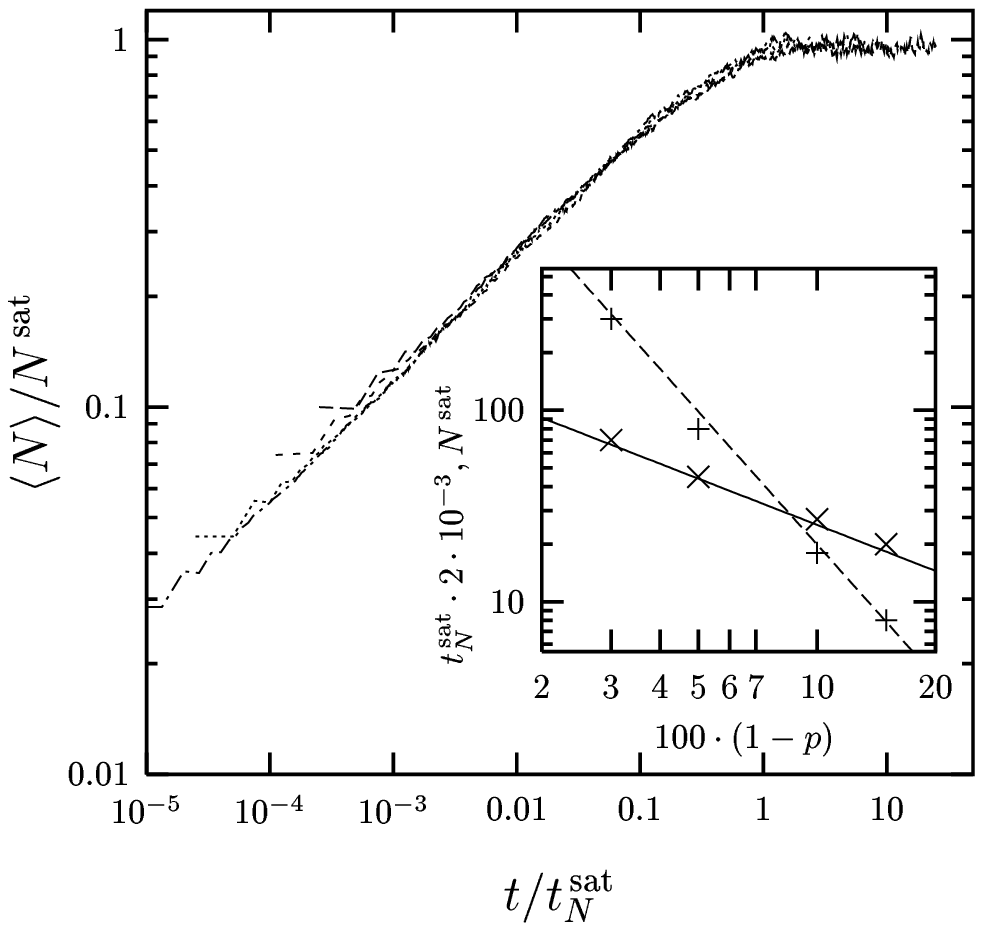}
\caption{Time evolution of the averaged number of nodes 
$\langle N \rangle$ 
(upper curves)
and the averaged mean connectivity 
$\langle\overline{k}\rangle$ (lower curves)
for different values of the parameter $p$. 
Values of $p$ from bottom to top are 
$p=0.85$, $0.9$, $0.95$, $0.97$, $1$ for number of nodes
and 
$p=0.85$, $0.9$, $0.95$, $0.97$ for connectivity.
Data are averaged over 1000 independent runs.
Straight lines are guides for eye with slopes 0.35
 (dashed)
and 0.24 (dot-dashed).
}
\label{fig:N-c-evol}
\caption{Data collapse of time evolution of the number of 
nodes $\langle N \rangle$ for different
values of the parameter $p$ from Fig. \ref{fig:N-c-evol}.
Inset:  Saturated values of the number of 
nodes $N^{\rm  sat}$ (crosses) 
and times of saturation $t^{\rm sat}_N$ (plusses) vs. $(1-p)$.
The straight lines are guides for eye with slopes 0.8 (full) and 2.3 (dashed).
}
\label{fig:N-collapse}
\end{figure}

We previously showed that the model is self-organized critical
 \cite{sla_ko_99,sla_ko_00}.
Here, we focus on scaling with time. 
In all simulations below, the evolution starts with a single node.

\section{Dynamic scaling of evolving network}
In this part, we restrict ourself to the situation $p_{\rm sing}=1$
(all isolated nodes will die out), but we consider different values of 
the parametr $p$.
The size  
$N$ as well as the connectivity $\overline{k}$ 
averaged over all nodes 
strongly fluctuate 
during the evolution (cf. Fig. 2 in \cite{sla_ko_99}),
and no clear tendency can be deduced. 
However,  two different regimes are identified
after averaging over many independent runs (Fig. \ref{fig:N-c-evol}).
A crossover time $t^{\rm sat}$
between two regimes depends on the parameter $p$.
In the initial stage $t\ll t^{\rm sat}$,
both averaged quantities 
increase as power-laws
\begin{equation}
\langle N\rangle \propto t^{\beta_N},  \hspace*{10mm}
\langle \overline{k}\rangle \propto t^{\beta_k}
\end{equation} 
with the exponents $\beta_N \simeq 0.35$ and
 $\beta_k \simeq 0.24$.
Here  $\langle ... \rangle$ is a statistical average,
and the bar denotes an average over nodes.
In the second regime, the averaged number of nodes $\langle N\rangle$  
and therefore also the averaged mean 
connectivity $\langle \overline{k}\rangle$ 
saturate after times $t^{\rm sat}_N$ and $t^{\rm sat}_k$.
The crossover times  $t^{\rm sat}_N$ and $t^{\rm sat}_k$
are different ($t^{\rm sat}_k$ being smaller that $t^{\rm sat}_N$).

\begin{figure}
\twofigures[scale=0.7]{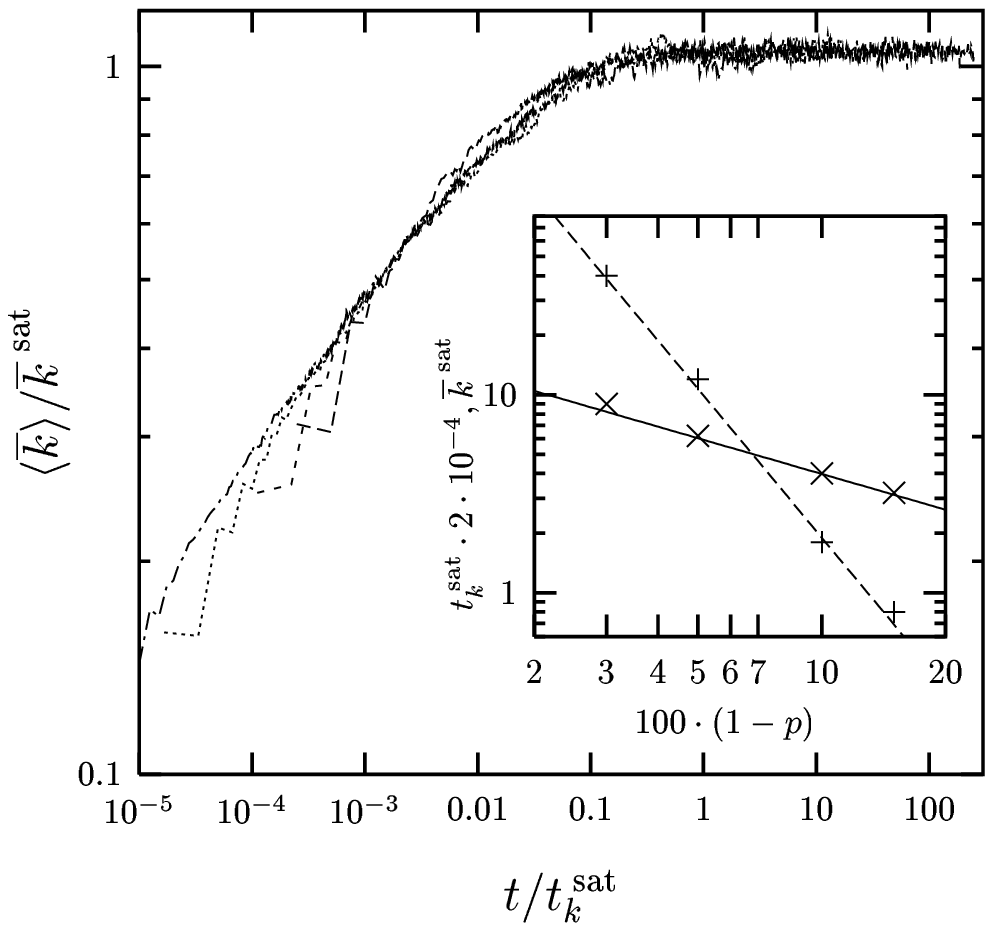}{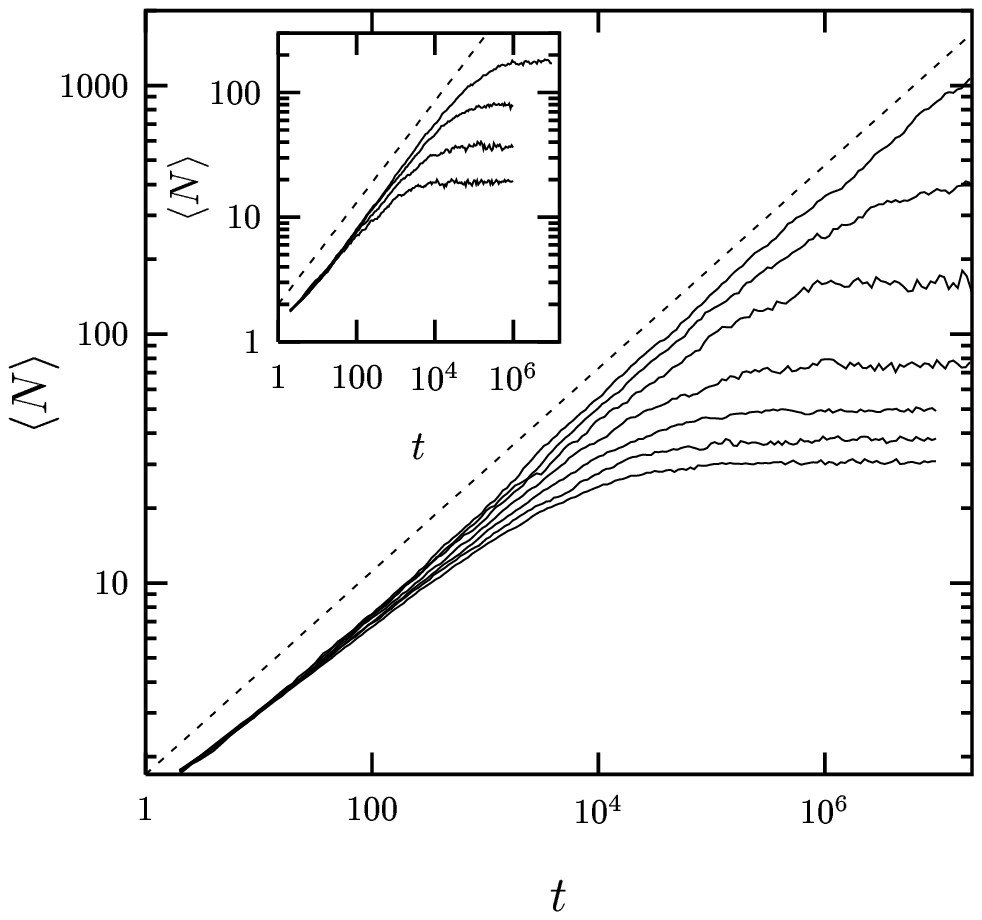}
\caption{Data collapse of time evolution of the 
averaged mean connectivity 
$\langle \overline{k}\rangle$ 
for different
values of parameter $p$ from Fig. \ref{fig:N-c-evol}.
Inset:  Saturated values of the averaged mean connectivity 
$\overline{k}^{\rm  sat}$ (crosses) 
and time for saturation $t^{\rm sat}_k$ (plusses) vs. $(1-p)$.
The straight lines are guides for eye with slopes 0.6 (full) and 2.5 (dashed).
}
\label{fig:c-collapse}
\caption{
Time evolution of the averaged number of nodes 
$\langle N \rangle$ 
in the lattice variant of the model with 8 neighbours
and 
for different values of the parameter $p_{\rm sing}$. 
Values of $p_{\rm sing}$ from top to bottom are $p_{\rm sing}=$ 
$0.05$, $0.1$, $0.2$, $0.4$, $0.6$, $0.8$,  $1$.
Data are averaged over 1000 to 10000 independent runs.
The straight line is guide for eye with slope 0.41.
Inset:
Time evolution of the averaged number of nodes 
$\langle N \rangle$ 
in the lattice variant of the model with 4 neighbours.
Values of $p_{\rm sing}$ from top to bottom are $p_{\rm sing}=$ 
$0.05$, $0.1$, $0.2$, $0.4$.
The straight line are guide for eye with slope 0.41.
}
\label{fig:NevolLat}
\end{figure}

The saturated values  $N^{\rm sat}$,  $\overline{k}^{\rm sat}$ and
the times of saturation  $t^{\rm sat}_N$ and $t^{\rm sat}_k$ increase 
with $p$.
The parameter $p$ controls correlations between 
the structure of the neighbourhood
of the ``mother'' node and the ``daughter'' node. 
We define $\xi =1/(1-p)$ as an analog of the correlation
length. We have found that the saturated values $N^{\rm sat}$,
$\overline{k}^{\rm sat}$  depend on $\xi$
as powers (insets in Figs. \ref{fig:N-collapse}
and \ref{fig:c-collapse})
\begin{equation}
N^{\rm sat} \propto \xi^{\alpha_N}\,,\hspace*{10mm}
\overline{k}^{\rm sat} \propto \xi^{\alpha_k}\,\hspace*{10mm}
\end{equation}
with the exponents 
$\alpha_N =0.8$, $\alpha_k =0.6$.
The exponents $\alpha_N$ and $\alpha_k$ approximately agree
with the exponents obtained in \cite{sla_ko_00}
by the calculation of the mean values using the stationary distribution 
${\cal P}(k)$.
Both crossover times also fulfil power-law 
\begin{equation}
t^{\rm sat}_N \propto \xi^{z_N}\,, \hspace*{10mm}
t^{\rm sat}_k \propto \xi^{z_k}\,. 
\end{equation}
The exponents are $z_N=2.3$ and $z_k=2.5$
(insets in Figs. \ref{fig:N-collapse} and \ref{fig:c-collapse}).
The behaviour  for  $p = 1$ is different.
There is no saturation for $p = 1$ on the scale of 
our simulations and the model is critical \cite{sla_ko_99,sla_ko_00}.

This scaling behaviour is similar to the well 
known dynamic scaling for kinetic roughening
during surface growth \cite{fam_vic_85}.
Data for $\langle N\rangle (t,\xi)$  as well as for 
$\langle \overline{k}\rangle (t,\xi)$ 
can be rescaled to a single curve (Figs. \ref{fig:N-collapse} 
and \ref{fig:c-collapse}).
Hence, both $\langle N\rangle$ and  $\langle \overline{k}\rangle$
fulfil the following scaling relations 
\begin{equation}
\langle N\rangle \propto \xi^{\alpha_N} f_N(t/\xi^{z_N}), \hspace{10mm}
\langle \overline{k}\rangle \propto \xi^{\alpha_k} f_k(t/\xi^{z_k}),
\end{equation}
where $f_N(x)$ and $f_k(x)$ are two different 
scaling functions with properties: 
$f_N(x) =  const_1$, $x \gg 1$,  $f_N(x) \propto x^{\beta_N}$, 
$x \ll 1$ and 
$f_k(x) = const_2$, $x \gg 1$,  $f_k(x) \propto x^{\beta_k}$, $x \ll 1$. 
Exponents $\beta_N$, $\alpha_N$, $z_N$ 
($\beta_k$, $\alpha_k$, $z_k$) are not independent but obey the relation
$\beta_N$$=$$\alpha_N/z_N$  
($\beta_k$$=$$\alpha_k/z_k$)
(for reviews on kinetic roughening see 
\cite{kr1,kr2}).
The relations between exponents are well satisfied for the 
above obtained values.

\section{Lattice variant of the model}
If $p_{\rm sing} < 1$ then
more and more isolated nodes will be generated.
The above rules do not allow to connect
an isolated node to any other node.
It will remain isolated forever.
This  may be the unrealistic situation.
One rather expects that isolated nodes can be with a certain probability
reconnected. In this case an additional rule for reconnection would be needed,
e.g., based on the list of links of extinct nodes.
This would imply the need of complicated book-keeping. 
In order to assess the effect of singular 
extinctions, we use a different
variant of the model which allows to avoid the above
unrealistic situation.

We suppose that nodes are now living on a regular lattice.
It means that only some sites of the given lattice are occupied by 
nodes at the given time. The occupation may change during evolution.
The lattice sites can represent the different positions in the 
Euclidean space or in some abstract space.
The  extremal dynamic is again emploied,
nodes are added or removed using the rule {\it (ii)}.
However, a node can speciate only if there is an unoccupied site 
in the  neighbourhood of the selected extremal node.
A new ``daughter'' node is placed on a randomly selected 
unoccupied site within the neighbourhood.

The structure of links of the ``daughter'' node
is now not inherited but it  is automatically given by
existing nodes in 
her neighbourhood. 
The links are established from a given node to  all 
neighbouring nodes (occupied sites within a given neighbourhood).
Hence, the above rule  {\it (iv)}
does not apply and is replaced be automatic connection within the given 
neighbourhood.
The network is formed by clusters of connected nodes 
on the lattice.
All  links of an extinct node are broken.
We apply the rule  {\it (vi)} 
for singular extinction with $p_{\rm sing}\,\epsilon\, <0,1>$. 
Therefore, isolated nodes can be created and can be reconnected.
The rules also allow reconnection of previously disconnected clusters.
The lattice variant of the model has similar self-organization properties
as the above described off-lattice variant. 
In particular, it exhibits power-law 
in the distribution of forward avalanches \cite{steiner00}.

We consider for simplicity in the following 
the simple square lattice of the size $L\times L$ with the periodic boundary
conditions. 
The size $L$ has been chosen so large that the finite size effects
 have no influence ($L$ = 40--160 depending on $p_{\rm sing}$).
We performed simulations for two definition of
 the neighbourhood.
The neighbourhood composed of 
 4 nearest neighbours (4n-model) and 
 4 nearest neighbours plus 4 next nearest neighbours (8n-model).

We found that the time behaviour of the network size and the 
averaged mean connectivity is qualitatively the same as in the off-lattice model.
Both quantities 
are fluctuating, however, after averaging (over 1000 or more independent
runs) two regimes can be clearly seen: the 
initial power-law increase and the saturation at the late times.
The time evolution of network size for 
two modifications with 4 and 8 neighbours is  
displayed in Fig. \ref{fig:NevolLat}.
The initial power-law increase 
$\langle N\rangle \propto t^{\beta_N} $
with exponent $\beta_N = 0.41$ is observed in both cases.
$N_{\rm sat}$ increases with decreasing $p_{\rm sing}$.  
Saturated values are substantially lower for 4n-model than for 8n-model.

\begin{figure}
\twofigures[scale=0.7]{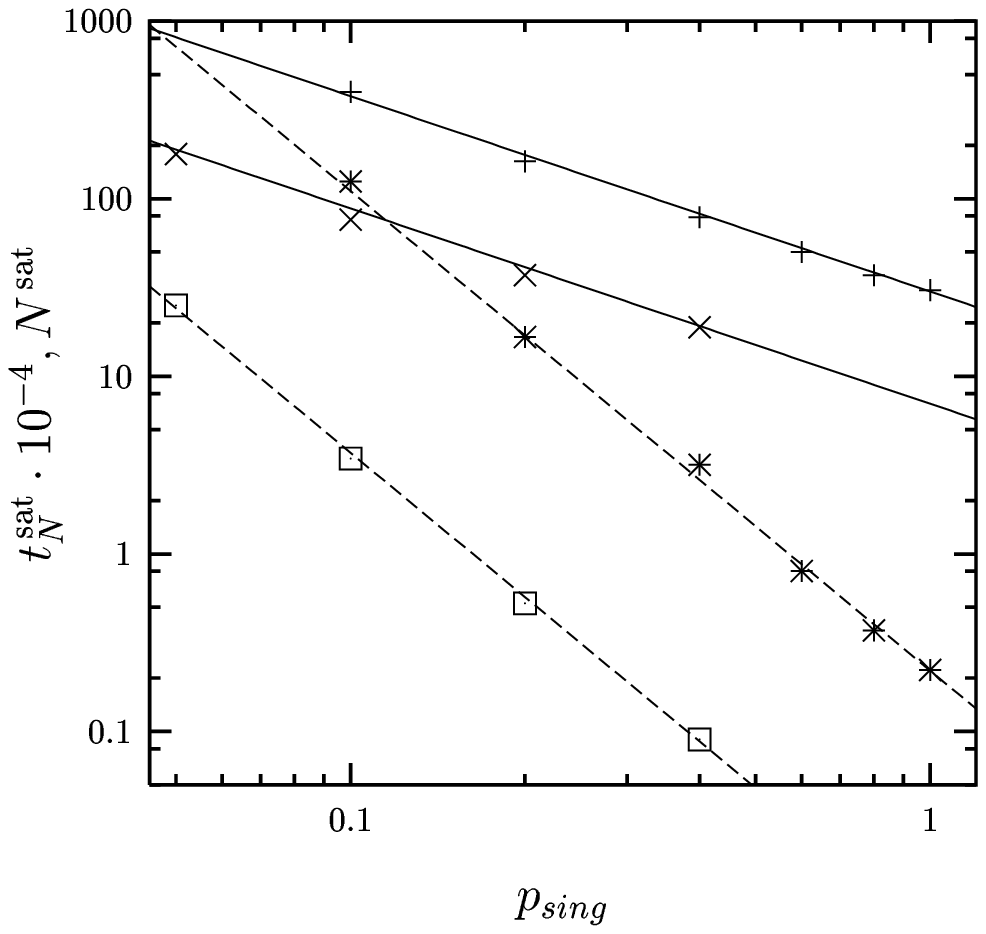}{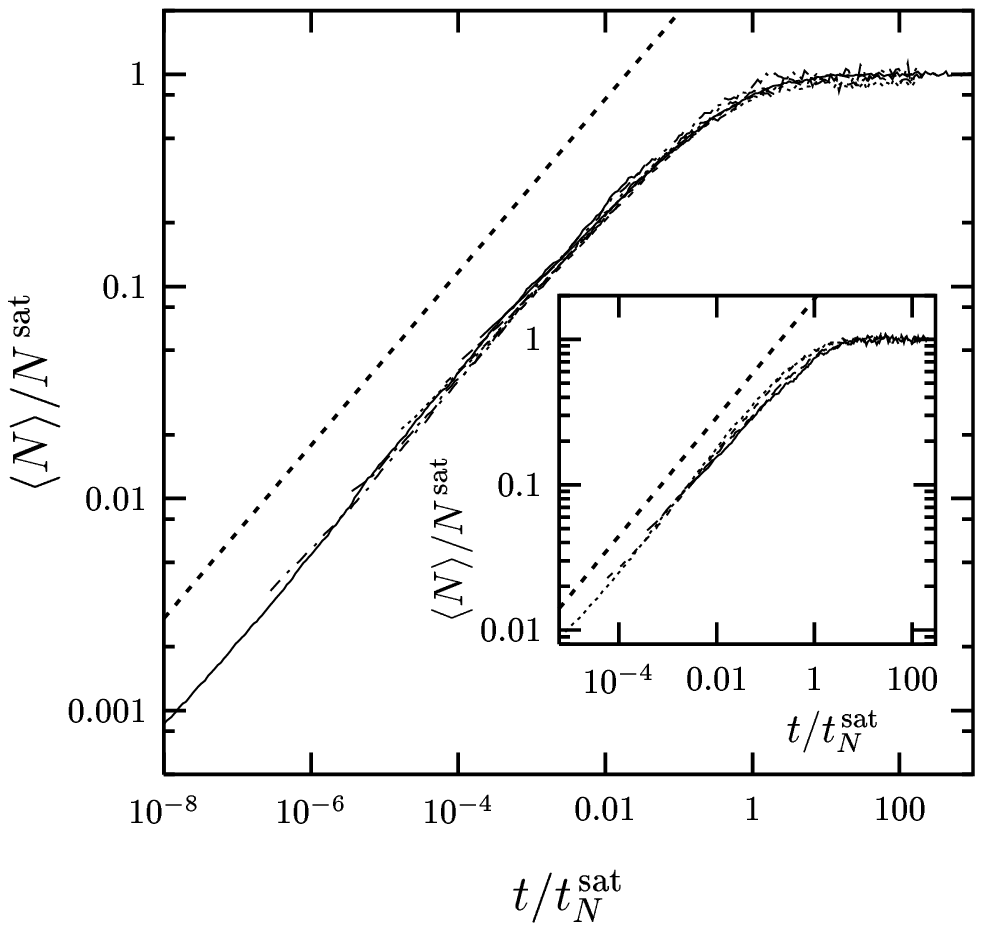}
\caption{Saturated values of the number of nodes
$N^{\rm sat}$ for the model with 4 neighbours (crosses)
and with 8 neighbours (plusses) and time of saturation
$t^{\rm sat}_N$
for the model with 4 neighbours (squares)
and with 8 neighbours (stars)
vs.  $p_{\rm sing}$.
The straight lines are guides for eye with slopes 1.1 (full) and 2.7 (dashed).
}
\label{fig:NsatLat}
\caption{Data collapse of time evolution of the number of 
nodes $\langle N \rangle$ in the lattice variant of the model
with 8 neighbours for different
values of parameter $p_{\rm sing}$ from Fig. \ref{fig:NevolLat}.
Inset:  Data collapse of time evolution of the number of 
nodes $\langle N \rangle$ in the lattice variant of the model
with 4 neighbours.
}
\label{fig:N-collapse2}
\end{figure}

The dependence of saturated values and times of saturation
on  $p_{\rm sing}$ is shown in Fig. \ref{fig:NsatLat}.
There are again power-laws  with more or less the same exponents
for both considered neighbourhoods.
Similarly as above, we define an analog of the correlation length
$\tilde{\xi} = 1/p_{\rm sing}.$
Then we have
\begin{equation}
N^{\rm sat} \propto \tilde{\xi}^{\alpha_N}\,,\hspace*{10mm}
t^{\rm sat}_N \propto \tilde{\xi}^{z_N}\,\hspace*{10mm}
\end{equation}
with the exponents 
 $\alpha_N = 1.1$ and $z_N = 2.7$.
The relation 
$\beta_N$$=$$\alpha_N/z_N$  is well satisfied.
We show the corresponding data collapse
supporting the scaling relation 
$
\langle N\rangle \propto \tilde{\xi}^{\alpha_N} f_N(t/\xi^{z_N})
$
in Fig. \ref{fig:N-collapse2}.
When $p_{\rm sing} = 0$ the network grows until the evolution will
be affected by the given finite lattice size.

Due to low maximal possible value, 
the connectivity saturates very soon
and good scaling of connectivity with
$p_{\rm sing}$ has not been obtained.
Therefore, data for time evolution of connectivity are not
presented.
In order to study carefully scaling of connectivity
in the lattice model, one should consider
a larger size of neighbourhood - include also more distant
sites or consider model on the hypercube.

An external condition limiting the size of the network or the connectivity
will lead trivially to the saturation of these quantities. One
expects that there will be  scaling with the limiting value.
We verified that indeed there is dynamic
scaling  
of the form
\begin{equation}
\langle N\rangle \propto L^{\alpha_N} g(t/L^{z_N})\, 
\end{equation}
with the time and the external parameter $L$.
Here, $g(x)$ is again 
scaling function with the properties: 
$g(x)=const_3$  for $x \gg 1$
$g(x)$$\propto$$x^{\beta_N}$ for
$x$$\ll$$1$. 
In the case of 4n-model,
we measured the exponents
$\alpha_N =2.07$, $\beta_N =0.44$, $z_N=4.8$
fulfilling well the relation $\beta_N$$=$$\alpha_N/z_N$.

\section{Conclusion}
We have studied the time dependence of global 
characteristics of evolving random networks.
We used the network model with 
the extremal dynamics and with the
variable rules for evolution of the local 
geometry.
We have found that the averaged network size and the averaged
mean connectivity
exhibit dynamic scaling with time and with
the parameter controlling the form of dynamical rules
which in turn affects spreading of the correlations during evolution.
We considered two different variant of the model:
the off-lattice variant with the varying reproduction 
of the local topology
and the lattice  variant with the variable amount of
removal of isolated nodes.
We have found that 
dynamic scaling is fulfiled in both variants of the model. 
However, the measured exponents are different.
We attribute this variance to the difference 
in mechanisms of evolution.
On the other hand,
we verified that there are universal features.
The exponents in the lattice variant  are 
robust to the change of 
the size of neighbourhood.

We expect that dynamic scaling can be observed also in other models
of evolving networks with the tunable internal dynamics.
In general, dynamic scaling of evolving networks 
can be characterized by four independent exponents, two for the
time dependence of the averaged size, and two for  the evolution 
of the averaged mean connectivity.
The exponents for evolution of the network size may be trivial 
or not defined in models in which the number of nodes
is only and permanently increasing.
Nevertheless, the connectivity may saturate and exhibit  nontrivial
scaling.
Varying the external parameters like the maximal size leads to a
qualitatively different scaling.
We believe that the study of dynamic scaling of evolving networks
might provide a clue to identification of 
universal features of network evolution.
It is an open question if dynamic scaling
will allow the classification of types of network evolution
into different universal classes
as in the case of the surface growth, or if evolution of global 
characteristics turns to be unique for
each individual network as it has been found for
exponents describing the distribution of connectivities in scale
free networks.

\acknowledgments
This work was supported by grant No. 202/01/1091 of the GA \v{C}R

\end{document}